\documentclass[pra,aps,twocolumn,showpacs,superscriptaddress,amsmath,amssymb]{revtex4-1}
\usepackage{color}

\usepackage{amsmath,graphicx}

\begin{document}
\def\tr{\rm{Tr}}
\def\la{{\langle}}
\def\ra{{\rangle}}
\def\a{{\alpha}}
\def\e{\epsilon}
\def\q{\quad}
\def\w{\tilde{W}}
\def\t{\tilde{t}}
\def\a{\hat{A}}
\def\h{\hat{H}}
\def\E{\mathcal{E}}
\def\p{\hat{P}}
\def\n{\phi_n}
\def\k{\phi_k}

\title{Zeno effect and ergodicity in finite-time quantum measurements}
%
%
\author {D. Sokolovski}
\affiliation{Departamento de Qu\'imica-F\'isica, Universidad del Pa\' is Vasco, UPV/EHU, Leioa, Spain}
\affiliation{IKERBASQUE, Basque Foundation for Science, E-48011 Bilbao, Spain}

\date{\today}
\begin{abstract}
We demonstrate that an attempt to measure a non-local in time quantity,  such as 
the time average $\la A\ra_T$ of a dynamical variable $A$, by separating Feynman paths into ever narrower 
exclusive classes  traps the system in eigensubspaces of the corresponding operator $\a$.
Conversely, in a long measurement of $\la A\ra_T$ to a finite accuracy, the system  
explores its Hilbert space and is driven to a universal steady state in which von Neumann ensemble average of $\a$ coincides with $\la A\ra_T$. Both effects are conveniently analysed
in terms of singularities and critical points of the corresponding amplitude distribution and the 
Zeno-like behaviour is shown to be a consequence of conservation of probability.


\end{abstract}

%
%
\maketitle
%
%
%
%
%
%
%
Quantum Zeno effect (see, for instance, \cite{Z0}-\cite{Z2} and Refs. therein)
is often associated with the perturbation frequent projective von Neumann measurements \cite{vN}
made on the observed system. Suppose, for example, that one wishes to determine the duration $\tau_{\Omega}$ a quantum system spends in a particular subspace $\Omega$ of its Hilbert space. Checking ever more frequently whether the system is indeed inside $\Omega$ one eventually destroys the transitions
between $\Omega$ and the rest of the Hilbert space. Thus, because of the Zeno effect,  
 a continuoulsy observed system prepared inside $\Omega$ would spend there all available time, while for a system initially outside
$\Omega$, $\tau_{\Omega}$ would be exactly zero \cite{ZW}.
Alternatively,
one can perform a {\it finite time} measurement of $\tau_{\Omega}$ in which a meter monitors
the system over a finite period of time, and one single observation is made at the end of the run.
 One example of such a meter is Larmor clock consisting of a spin which rotates only when the system resides in $\Omega$.  From the clock's final orientation one is able to determine the value of $\tau$, but can learn nothing about the precise moments the system enters and leaves $\Omega$. A conceptually similar measurement of the duration $\tau_{q}$ a qubit spends in its state $|q\ra$  proposed in \cite{S1} employs a large number of bosons (e.g, a weakly interacting Bose-Einstein condensate) trapped in one of the wells of a symmetric double well potential. With atomic current between the wells increased whenever the qubit occupies the state of interest, the number of bosons found in the other well contains, like a reading of a Larmor clock, information about $\tau(q)$.
  The purpose of this paper is twofold: we analyse the Zeno effect arising in  the high accuracy limit of a general finite-time measurement. We also study
the effects of a measurement 
of long-time average of a quantum variable and search for any evidence of ergodic behaviour.
 Various approaches to quantum ergodicity can be found in Refs.  \cite{E1} - \cite{E3}, with the importance of measurement(s) performed on the system emphasised in \cite{E2}. 
 A variant of the Zeno effect arising solely from strong interaction between a system and its environment has been studied in  \cite{Z2}.

Consider a quantum system in an $N$-dimensional Hilbert space with a Hamiltonian $\h$.
Choosing an orthogonal basis $|a_m\ra$, $m=1,2,..N$ in which $\h$ is not diagonal, we 
can write a transition amplitude between initial and final states $|i\ra$ and $|f\ra$ over a time $T$ as a sum over Feynman paths

\begin{eqnarray}\label{1} \nonumber
U^{f \leftarrow i}(T) \equiv \la f|\exp(-i\h T|i\ra = 
lim_{K\rightarrow \infty}\sum_{m_1,..,m_K}^N  \la f|
a_{m_K}\ra \times\\
\prod_{j=1}^{K-1}\la a_{m_{j+1}}|\exp(-i\h \e |a_{m_{j}}\ra\la a_{m_1}|i\ra 
\equiv \sum_{paths} U^{f \leftarrow i}[path],\quad  \quad
\end{eqnarray}
where $\e\equiv T/(K-1)$ and each path is defined by a sequence $\{ m_1,...,m_K\}$
numbering the states $|a_m\ra$ through which the system passes until reaching
the final state $|f\ra$.
Consider a quantity $A$ represented by an operator $\a$  diagonal in the chosen representation,
\begin{eqnarray}\label{1a} 
\a=\sum_n A_n \p_n,\quad \p_n\p_m=\p_n\delta_{mn}, \quad \sum_n \p_n=1,
\end{eqnarray}
where $\p_n$ are mutually orthogonal projectors on (one- or multidimensional) subspaces spanned by
vectors $|a_{m}\ra$ corresponding to the same eigenvalue $A_n$.
We wish to measure the value of a Feynman functional
\begin{eqnarray}\label{2} 
F[path] = lim_{K\rightarrow \infty}\sum_{j=1}^{K-1}\beta (j\e)A_{m_j}\e \\
 \equiv \int_0^T \beta(t)A(t)dt
 \nonumber 
\end{eqnarray}
where $\beta(t)$ is a known function \cite{FOOT1} and $A(t)$ denotes the (highly irregular)
function traced by the value of $A$ along a given Feynman path. 
Equation (\ref{2}) may,
for example, represent the time average of a quantity $A$ if one chooses $\beta(t)=1/T=const$ and,
in particular, the fraction of the time, $\tau_{\Omega}/T$,  the system has spent in 
$\Omega$ if $\a$ is also chosen to be the projector onto a subspace $\Omega$ spanned by $|a_n\ra$,
$\a=\p_{\Omega} \equiv \sum_{n \in \Omega}|a_n\ra\la a_n|$
\cite{S1}.
The probability amplitude for the value of $F$ to be $y$ is
 given by the restricted path sum 
\begin{eqnarray}\label{3} 
\Phi^{f \leftarrow i}(y,T)= \sum_{paths}\delta(F[path]-y) U^{f \leftarrow i}[path].
\end{eqnarray}
where $\delta(z)$ is the Dirac delta.
Without loss of generality we choose the measuring device to be a
Neumann pointer with position $y$ which interacts with the system over a time $T$,
 the full Hamiltonian being $\hat{\mathcal{H}}(t) = \h-i\partial _y \beta(t) \a$.
At $t=T$ the pointer states of the meter, $|y\ra$, are entangled with the 
system's states obtained by propagation along Feynman paths satisfying the condition 
$F[path]=y$ \cite{S2}. In particular, for the system and the meter prepared at $t=0$  in an product state
$|i\ra |G\ra$, 
the probability amplitude to find at $t=T$ the system in the state $|f\ra$ and,
simultaneously, the pointer reading $y$, $\Psi^{f \leftarrow i}(y)$, is given by 
\begin{eqnarray}\label{4} 
\nonumber
  \Psi^{f \leftarrow i}(y,T)\equiv \la y|\la f|\exp[-i\int_0^T \hat{\mathcal{H}}(t)dt]|i\ra|G\ra = \\
\int G(y-y') \Phi^{f \leftarrow i}(y',T) dy,
\end{eqnarray}
where $G(y) \equiv \la y|G\ra$. Note that the second of Eqs.(\ref{4}) is not specific to our choice of von Neumann meter and occurs for a wider class of quantum measurements, e.g., for the one considered in \cite{S1}.
With $G(y)$ narrowly peaked around the origin, Feynman paths with different values
of $F[path]$ contribute to different final meter states,
 and $| \Psi^{f \leftarrow i}(y)|^2$ yields the probability to reach the final state and obtain the value $y$ for the quantity in Eq.(\ref{2}). The measured result contains, however, an intrinsic quantum uncertainty as the values of $y'$ within the peak's width around $y$ remain indistinguishable. With many different Feynman paths contributing to the transition (\ref{1}) one might expect the values of $F$ to have a broad distribution, which a more accurate measurement would resolve in ever greater detail. 
The accuracy can be improved by making the initial pointer state narrower in the coordinate space, e.g., by replacing $G(y)$ with
\begin{eqnarray}\label{5} 
G_{\alpha}(y)= \alpha^{1/2} G(\alpha y), \q\q lim_{\alpha \rightarrow \infty} |G_{\alpha}(y)|^2=\delta(y)
\end{eqnarray}
where the factor $\alpha^{1/2}$ ensures the correct normalisation of the new state, $\int  |G_{\alpha}(y)|^2 dy = \int |G(y)|^2 dy =1$ .  

With the help of Eqs.(\ref{4}) and (\ref{5}) we can now prove a general result: 
an accurate finite time measurement,
$\alpha \rightarrow \infty$,  $T<\infty$,
would indicate that at all times $A(t)$ maintains a constant value equal to one of the eigenvalues $A_n$,
with the value of the functional (\ref{2}) 
equal exactly to $A_n\int_0^T \beta(t)dt$.
The proof follows from observing first that $\Phi^{f \leftarrow i}(y)$ cannot be a smooth function for all final states $|f\ra$. Indeed, increasing $\alpha$
while maintaining unit normalisation will cause the integral $\int G_{\alpha}(y)dy = \alpha^{-1/2}
\int G(y)dy$ to vanish. With it would also vanish $\Psi^{f \leftarrow i}(y)$, $\lim_{\alpha \rightarrow \infty}\Psi^{f \leftarrow i}(y) = \Phi^{f \leftarrow i}(y)\alpha^{-1/2}\int G(y-y')dy' = 0$, thus
contradicting conservation of probability for the pointer.  
Thus, $\Phi^{f \leftarrow i}(y)$ must have a singular part, which we evaluate
by rewriting Eq.(\ref{3}) as a Fourier integral
\begin{eqnarray}\label{6} 
\Phi^{f \leftarrow i}(y)=\q\q\q\\
\nonumber  (2\pi)^{-1}\int \exp(i\lambda y) \la f|\exp[-i\int_0^T (\h + \lambda \beta(t)\a)dt]|i\ra d\lambda.
\end{eqnarray}
Further, writing $\exp\{-i\int_0^T [\h + \lambda \beta(t)\a]dt\} = lim_{K\rightarrow \infty}
\prod_{j=1} ^K\exp\{-i[\h + \lambda \beta(jT/K)\a]T/K\}$ we note that
$lim_{\lambda \rightarrow \infty}\exp\{-i[\h + \lambda \beta(jT/K)\a]T/K\}=
\sum_n  \exp\{-i\lambda \beta(jT/K) A_n\}\exp(-i\p_n\h\p_n T/K)\p_n  + O(\lambda^{-1})$.
Defining a Zeno hamiltonian as (c.f. \cite{Z2})
\begin{eqnarray}\label{6a} 
\h_Z=\sum_n \p_n\h\p_n, 
\end{eqnarray}
we obtain 
\begin{eqnarray}\label{7} 
 \la f|\exp[-i\int_0^T (\h + \lambda \beta(t)\a)dt]|i\ra = \q\q\q\\
 \nonumber
 \sum_n \exp\{-i\lambda A_n\int_0^T\beta(t)dt\} \la f|\exp(-i\h_Z T)\p_n|i\ra\\
  \nonumber
  +\la f|\hat{u}(\lambda)|i\ra
\end{eqnarray}
where the last terms vanishes as $\lambda \rightarrow \pm \infty$, $lim_{\lambda \rightarrow \infty}\la f|\hat{u}(\lambda)|i\ra=O(\lambda^{-1})$. 
 Inserting Eq.(\ref{7}) into Eq.(\ref{6}) yields
\begin{eqnarray}\label{8} 
\Phi^{f \leftarrow i}(y)=
\sum_n\la f|\exp(-i\h_ZT)\p_n|i\ra\times \\
 \nonumber  \delta (y-A_n\int_0^T\beta(t)dt) + \Phi^{f \leftarrow i}_{smooth}(y).
\end{eqnarray}
where
 $\Phi^{f \leftarrow i}_{smooth}(y)$
  is the smooth  Fourier transform of the last term in Eq.(\ref{7}). 
With the contributions from the smooth term vanishing in the limit $\alpha \rightarrow \infty$ 
\cite {FOOT0}  
  the monitored system is seen to undergo unitary 
  evolution with a reduced Hamiltonian $\p_n\h\p_n$ in the subspaces corresponding 
  to each of the distinct eigenvalues $A_n$, $n=1,2..$. As in the case of the Zeno effect caused by frequent observations,\cite{Z0}-\cite{Z2},  an accurate finite-time measurement suppresses transitions between different subspaces. 
 Taking trace over the system's variables we find the probability distribution for 
 the functional (\ref{2})
\begin{eqnarray}\label{9} 
\nonumber
lim_{\alpha \rightarrow \infty }W^i (y,T) \equiv \sum_f |\Phi^{f \leftarrow i}(y,T)|^2 \approx \\
\sum_n \delta(y-A_n\int_0^T\beta(t)dt)\la i|\p_n|i\ra.
\end{eqnarray}
In particular, a strongly observed system starting in a a state corresponding to  a non-degenerate eigenvalue,
 $|i\ra=|a_M\ra$, $\p_M=|a_M\ra \la a_M|$,
 would follow a constant Feynman path $A(t)=A_M$, thus having  
 the time average of $\a$ exactly equal to $A_M$, and spending
all available time 
in its initial state, $\tau_M=T$. 
This failure to find real evidence of the irregular virtual
motion suggested in Eq.(\ref{1}) by separating Feynman paths into ever narrower 
exclusive classes according to the value of a functional (\ref{2}) constitutes the finite-time Zeno effect and is the first result of this paper.

Next we show that the evidence of the virtual motion is recovered
if one performes a long measurement of an arbitrarily high but fixed accuracy, $\alpha < \infty$ and $T\rightarrow \infty$.
For simplicity we consider the time average of $A$, thus choosing $\beta(t)=1/T=const$.
Changing variables in Eq.(\ref{6}), $z=\lambda/T$, and using spectral representation for the 
operator $\h + z \a $ yields
\begin{eqnarray}\label{10} 
\Phi^{f \leftarrow i}(y)=(2\pi)^{-1}T\times \q\q\q\\ \sum_{n=1}^N
\nonumber  \int \la f| \psi_n(z) \ra\exp\{i[zy-\E_n(z)]T\}\la  \psi_n(z) |i\ra dz.
\end{eqnarray}
where $(\h + z \a)| \psi_n(z) \ra=\E_n(z)| \psi_n(z) \ra$. The long time behaviour of 
$\Phi^{f \leftarrow i}(y)$ is now determined by the critical points $z_n^s(y)$ of the exponent in Eq.(\ref{10}),
\begin{eqnarray}\label{11} 
\partial_z \E_n(z)_{z=z_n^s}=y.
\end{eqnarray}
Evaluating the integrals in Eq.(\ref{10}) by the stationary phase method yields $N$ rapidly oscillating contributions containing factors $\exp[iS_n(y)T]$ with the phases given by the Legendre transforms of $\E_n(z)$,
$S_n(y)=z_n^s y-\E_n(z_n^s)$. 
The critical points $y_n^s$ of $S_n(y)$ are determined by the condition $z_n^s(y_n^s)=0$ so that
from Eq.(\ref{11}) we have $y_n^s=\partial_z \E_n(z)|_{z=0}$. Calculating the derivatives 
with the help of the perturbation theory  and evaluating the integrals in Eq.(\ref{4}) by the stationary
phase method,
 we find 
\begin{eqnarray}\label{12} 
\nonumber
lim_{T\rightarrow \infty }\Psi^{f \leftarrow i}(y)=\q\q \\
 \sum_n G_\alpha(y-\la \n|A|\n\ra)\la f|\n\ra \exp(-i E_n T)\la \n|i\ra, 
\end{eqnarray}
where $\h |\n\ra=E_n|\n\ra$.
Extension to mixed states is straightforward.
For an initial state $\hat{\rho}_i$, all $E_n$ non-degenerate and all  $\la \n|A|\n\ra$ distinct, 
the probability distribution of the meter's readings is given by
\begin{eqnarray}\label{13} 
lim_{\alpha \rightarrow \infty }lim_{T\rightarrow \infty } W^i (y,T) = \\
\nonumber
\sum_n \delta(y-\la \n|\a|\n\ra)\la \n|\hat{\rho}_i|\n\ra,
\end{eqnarray}
where the order in which the limits are taken is essential. 
 \begin{figure}[h]
\includegraphics[width=6.5cm, angle=0]{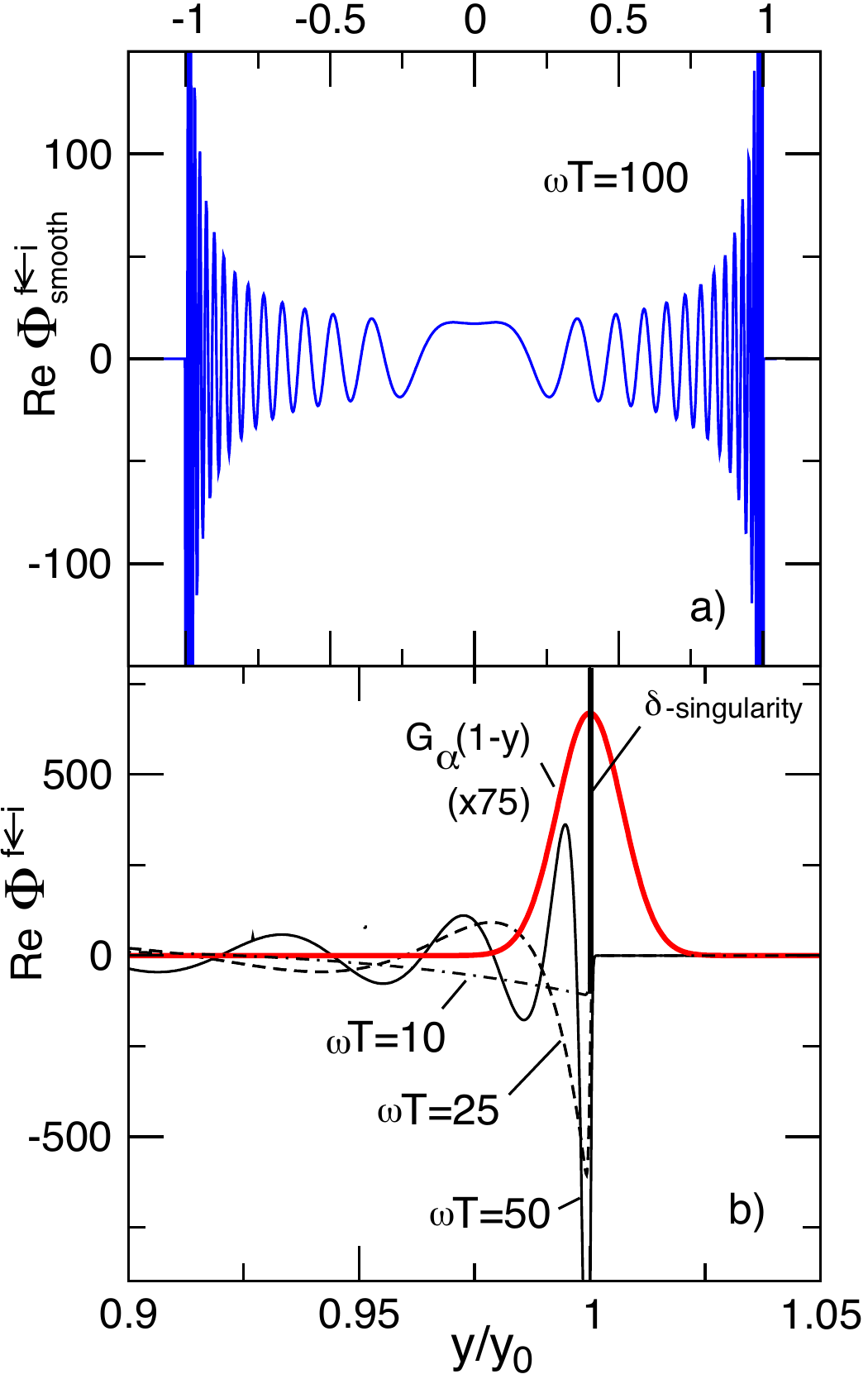}
\caption{(colour online) a) Amplitude distribution $ \Phi^{f \leftarrow i}_{smooth}(y,T)$ for 
$|i\ra=|1\ra$, $|f\ra=|+\ra$, $\h=\omega \hat{\sigma}_z$, $\a=\hat{\sigma}_y$, and $\omega T=100$; 
b) $ \Phi^{f \leftarrow i}_{smooth}(y,T)$ for $\omega T=50$ (solid), $\omega T=25$ (dashed),
and $\omega T=10$ (dot-dashed). Also shown is $G_{\alpha}(1-y)$  ($\times 75$) for $\alpha=100$ (thick solid).} 
\label{FIG3}
\end{figure}
We note that a strongly observed system prepared with a known energy $E_{k}$, $\la \phi_n|\hat{\rho}_i|\n\ra=\delta_{kn}$,
explores its Hilbert space is such a way that  the long-time 
time average of a dynamical variable $A$ is sharply defined, with the value equal to the ensemble average in the projection von Neumann measurement of the operator $\a$, $\la \k|\a|\k\ra$.
In particular, as $T\rightarrow \infty$, the fraction of time a system prepared in a pure state $|\k\ra$ spends in a subspace $\Omega$ spanned by the 
subset of
 eigenvectors \{$|a_m\ra, m\in \Omega\}$ tends to
the measure of the subset $\mu_\Omega = \sum_{m \in \Omega}|\la a_m|\k\ra|^2$
This ergodic-like \cite{FOOTN} property of bound quantum motion, to our knowledge not yet discussed in literature,
is the second result of this paper.
Further, as seen from Eq.(\ref{12}),  a system starting in a mixed state $\hat{\rho}_i$,
undergoes relaxation to the same steady state diagonal in the energy representation,
 $\hat{\rho}_{f}=
\sum_n|\n\ra\la \n|\hat{\rho}_i|\n\ra\la \n|$, regardless of the choice of $\a$. Thus,  finite-time average  of a quantity $A$, 
$\la A\ra_T \equiv \int y W^i(y,T)dy$, tends, as $T\rightarrow \infty$, to the von Neumann ensemble
average $Tr\{ \a \hat{\rho}_{f}\} $. As in the case of a frequently observed system \cite{E3} the equivalence between the time- and ensemble averages is established in the state produced at the end of 
measurement,
$ \hat{\rho}_{f}$, rather that in the initial state  $ \hat{\rho}_{i}$.

We illustrate the above with a simple example, where one wishes to measure the time average 
of the $y$-component of a spin $1/2$, $\a \equiv \hat{\sigma}_y$, for a  two-level
system with Hamiltonian $\h=\omega  \hat{\sigma}_x$ prepared in its eigenstate $|1\ra$,
$\h|1,2\ra=\pm\omega |1,2\ra$. Choosing in Eq.(\ref{2})  $\beta=const$ we then convert to dimensionless 
variables, $T\rightarrow \omega T$, $y\rightarrow y/y_0$, $y_0\equiv \beta T$, $\beta \rightarrow 1/T$.
As in Ref.\cite{S1},  $G(y)$ in Eq.(\ref{4}) is chosen to be a Gaussian,
\begin{eqnarray}\label{14} 
G_{\alpha}(y)= (2/\pi)^{1/4}(\alpha)^{1/2} \exp(-\alpha^2y^2).
\end{eqnarray}
Figure 1a shows $ \Phi^{+\leftarrow 1}_{smooth}(y)$,
$\hat{\sigma}_z|\pm\ra = \pm |\pm\ra$,
 for $\omega T =100$, with the stationary
region clearly seen around $y_2^s=y_1^s= \la 1|\hat{\sigma}_y|1\ra=0$.
 \begin{figure}[h]
\includegraphics[width=8.5cm, angle=0]{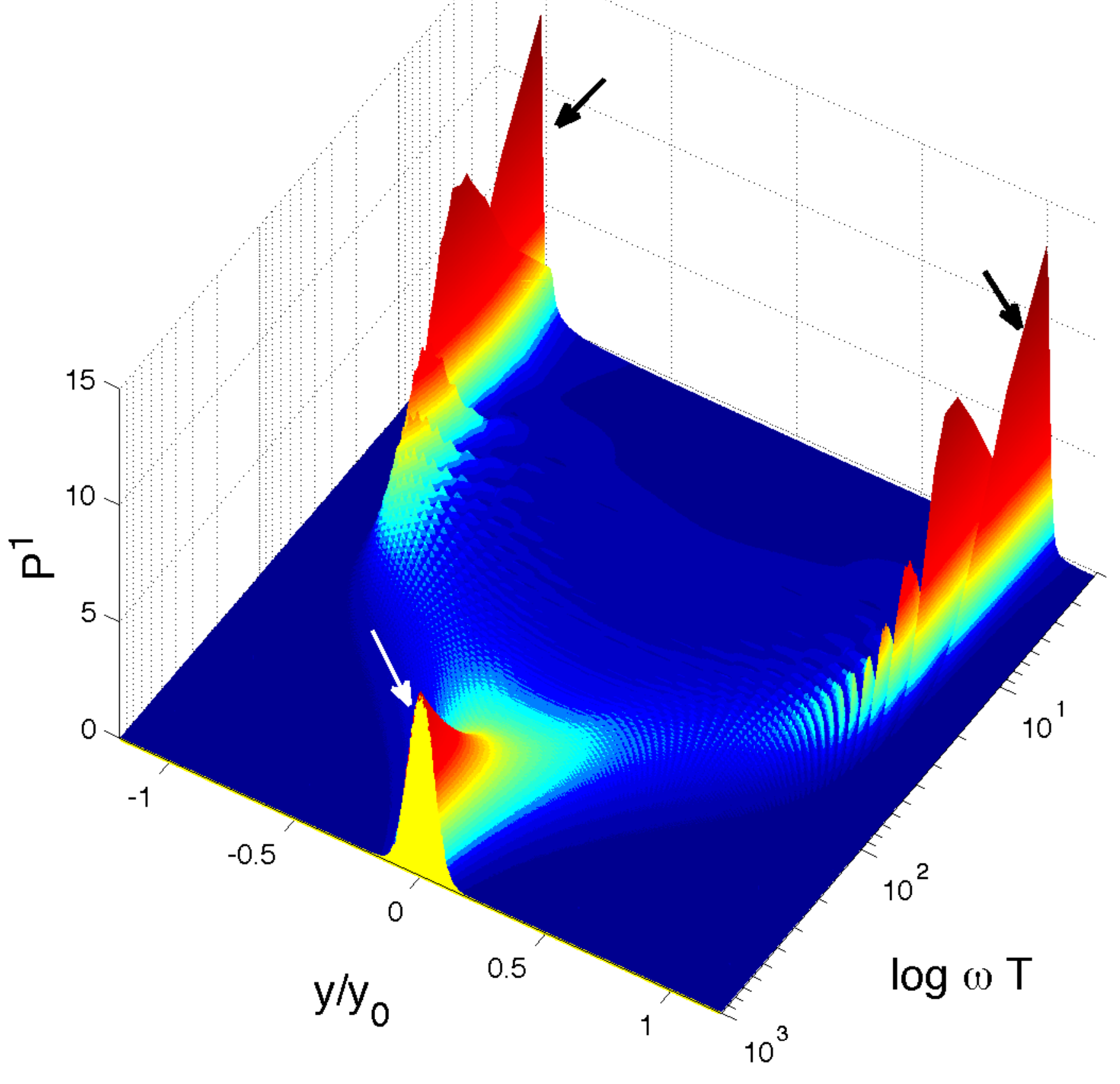}
\caption{(colour online) Probability distribution of the time average
of the $y$-component of the spin, $\a=\hat{\sigma}_y$, for a two-level system with the Hamiltonian $
\h=\omega \hat{\sigma}_x$ vs. the averaging time $T$. The system is prepared in the eigenstate $|1\ra$,
$\h|1\ra=\omega |1\ra$, and the accuracy of the measurement is $\Delta y/y_0=0.05$. The ergodic peak and  two Zeno peaks are indicated by white and black arrows, respectively.}
\label{FIG3}
\end{figure}
Two Zeno peaks at  $y/y_0=\pm 1$ predicted by Eq.(\ref{9}) and a single 'ergodic' (for want of a better word) peak predicted by Eq.(\ref{13}) are shown in Fig.2.
Transition between the regimes described by Eqs.(\ref{9}) and (\ref{13})
occurs as the Zeno peaks, however narrow $G_\alpha$ may be,
are eventually cancelled by the contributions from $ \Phi^{f \leftarrow i}_{smooth}(y)$,
whose oscillations become more rapid as the time $T$ increases. 
This is illustrated in Fig. 1b, for the spin-$1/2$ system described above, with $ \Psi^{+\leftarrow 1}(y=1)$ given by Eq.(\ref{4}). For $\omega T=10$ the contribution from $ \Phi^{+\leftarrow 1}_{smooth}(y)$ (dot-dashed) is negligible, and the Zeno peak is formed by the $\delta$-singularity of $ \Phi^{+\leftarrow 1}$ (c.f. Eq. (\ref{8})) at $y=1$. For $\omega T=50$ the contribution of the singular term is largely cancelled by the first negative oscillation of $ \Phi^{+\leftarrow 1}_{smooth}(y)$ (solid) that fits under the Gaussian
$G_\alpha(1-y)$ (thick solid), thus making $ \Psi^{+\leftarrow 1}(y=1)$ negligible.

In summary, we have considered a general finite-time measurement based on separating Feynman paths into exclusive classes according to the value of a functional such as time average of a dynamical variable
$A$, $\la A\ra_T$. We have shown that:
(i) A highly accurate measurement of a fixed duration
traps the measured system in the eigenstates (eigensubspaces) of the corresponding 
operator $\a$. 
(ii) For any quantity $A$, a  prolonged measurement  of an arbitrary but fixed accuracy destroys coherences in the energy representation, thus leaving  the
system in a steady state with $\la A\ra_T$ equal to the von Neumann  projection average of $\a$.
In the special case of a system prepared in a pure stationary state, $\la A\ra_T$ is sharply defined, 
and the proportion of time spent in a given subspace $\Omega$ is exactly equal
to the von Neumann probability to find the system there.
Both effects are readily explained in terms of singularities and critical points of the corresponding amplitude distribution. Present analysis can be extended to the cases of 'self-measurement', such as  wavepacket tunnelling \cite{S3}, where no external meter is employed.

This work was supported by the Basque Goverment grant IT472
and MICINN (Ministerio de Ciencia e Innovaci—n) grant FIS2009-12773-C02-01.


\begin{thebibliography}{999}
\bibitem{FOOT1} We assume $\beta(t)\ne 0$ in every finite interval of $[0,T]$.
\bibitem{FOOT0}  
More precisely, by Parceval's theorem $ \Phi^{f \leftarrow i}_{smooth}(y)$ is a square
integrable function, $\int |\Phi^{f \leftarrow i}_{smooth}(y)|^2dy <\infty$. Therefore
$|\Phi^{f \leftarrow i}_{smooth}(y)|^2$ can have, at most, integrable singularities which
would be even weaker for $|\Phi^{f \leftarrow i}_{smooth}(y)|$. Thus the integrals 
$\int G_\alpha (y-y')\Phi^{f \leftarrow i}_{smooth}(y')dy'$ are finite and would vanish 
for $\alpha \rightarrow \infty$.
\bibitem{FOOTN}
 We do not address the macroscopic aspect central to most studies of quantum ergodicity
 \cite{E1}, \cite{E2}. 
 We do, however, address non-trivial statistical behaviour of an
 observed completely specified quantum system, something not considered 
if an unperturbed Schroedinger evolution is assumed \cite {E3}.
\end{thebibliography}
\end{document}